# EVIDENCE FOR RECENT ACCRETION IN NEARBY GALAXIES

Dennis Zaritsky

UCO/Lick Observatory and Board of Astronomy and Astrophysics, Univ. of California, Santa Cruz, CA, 95064, E-mail: dennis@lick.ucsc.edu

## ABSTRACT

I discuss observations of magnitude residuals from the B-band Tully-Fisher relationship, $B - V$ color, chemical abundance gradients, and asymmetries in the H I and stellar disks of nearby spiral galaxies within the context of a model in which small satellites or H I clouds are accreted onto the outer disks of spiral galaxies. Correlations between the various observables support the hypothesis that accretion dilutes the gas phase abundances in the outer disk, steepens the abundance gradient across the disk, increases the star formation rate, and creates asymmetries in the outer disk. By estimating the duration of steep abundance gradients, elevated rates of star formation, or outer disk asymmetries, constraints can be placed on the rate of accretion events. The data suggest that accretion events at the current time are common.

## 1. INTRODUCTION

Although the frequency of galaxy interactions is poorly constrained, such events do occur (cf. Arp 1966) and they affect galaxy evolution (Lonsdale, Persson, & Matthews 1984; Kennicutt *et al.* 1987). Although major mergers are spectacular, the more common interaction is presumably the accretion of a small companion galaxy or gas cloud by a larger galaxy. If such interactions can be detected for some time after the event, one could measure the accretion rate and study its effect on the evolution of normal galaxies.

Which observables might attest to recent accretion in nearby spirals? By analogy to major mergers (Lonsdale, Persson, & Matthews 1984; Kennicutt *et al.* 1987; Mihos, Richstone, & Bothun 1992), accretion events might elevate the star formation rate (SFR). If so, it may be possible to identify galaxies that have recently accreted by their H$\alpha$ fluxes, colors, or absolute blue magnitudes, $M_B$. A second possible result of accretion, one which has often been invoked in models of Galactic chemical evolution (cf. Sommer-Larsen 1991 and references therein), is the dilution of existing metal enriched gas by infalling low-metallicity gas. Because satellite orbits decay faster into the disk plane than they do in radius (Quinn & Goodman 1986), gas-phase chemical abundance anomalies in the *outer disk* may be a further sign of recent accretion. Lastly, one might expect even minor interactions to affect a galaxy's morphology (cf. Schweizer *et al.* 1990).

In this *Letter* I discuss why the magnitude residuals from the standard B-band Tully-Fisher relationship, $B - V$ color, the slope of the O/H abundance gradient, and disk asymmetries may be indicators of recent accretion. I show that the relationships among these observables for a sample of nearby galaxies are qualitatively consistent with the predictions of the accretion model.

## 2. DATA AND ANALYSIS

The data used in this study are from published studies, with the exception of the chemical (O/H) abundances gradients, $G$'s, for NGC 2336 and NGC 4939 which come from Zaritsky & Kennicutt (1995) and are $-1.01 \pm 0.36$ and $-0.59 \pm 0.16$ dex/isophotal radius. All other $G$'s and the associated uncertainties are adopted as given by Zaritsky, Kennicutt, & Huchra (1994; hereafter ZKH) except for NGC 3521. I reject the second outermost H II region in NGC 3521 (see ZKH Fig. 8) as an outlier and measure $G = -0.83 \pm 0.34$. Only unbarred galaxies are included in Table 1 because bars affect abundance gradients (cf. Pagel *et al.* 1979; Vila-Costa & Edmunds 1992, ZKH, Martin & Roy 1994). Inclinations, $i$, and H I linewidths are adopted from Huchtmeier and Richter (1989) and references therein. The sample is confined to galaxies with high inclinations, $i > 30°$, with the exception of M101 which was included because it is one of the few galaxies with known Cepheids. H I linewidths are corrected for inclination and internal velocity dispersion as described by Tully (1988). Uncertainties of 5% in corrected H I linewidth, $W_R^i$, and 5° in $i$ are adopted. The $1\sigma$ distance uncertainties are set by the respective authors for the Cepheid and maser distances, to 20% for NGC 253 and NGC 7793 (which are placed in the Sculptor group at the distance of NGC 300), and to 30% for the other galaxies. Table notes provide additional information and references.

The presence of young (age < 1 Gyr) stars can be ascertained using either a luminosity normalized color, such as $B - V$, or a mass normalized color, such as that obtained by comparing $M_B$ for galaxies of similar mass. $B - V$ for galaxies correlates strongly with the H$\alpha$ emission (Kennicutt & Kent 1983) and so is an indicator of the SFR. A principal advantage of $B - V$ is that it is distance independent (for low redshift where K-corrections are negligible). Its greatest disadvantage is its dependence on the luminosity and mass of the galaxy (cf. Roberts & Haynes 1994). The mass normalized color bypasses this weakness, but is instead distance dependent. I use the Tully-Fisher relationship (Tully & Fisher 1977; hereafter TF) to calculate a galaxy's mass-normalized color. The difference between the observed $M_B$ and the predicted $M_B$ is the B-band TF residual magnitude (hereafter referred to as $\Delta B$).

To calculate $\Delta B$'s, I adopt the B-band TF relationship presented by Pierce & Tully (1992); $M_B = -19.55 - 7.48(\log W_R^i - 2.5)$. The residuals from this line (cf. Table

TABLE 1
NEARBY GALAXY SAMPLE

| NGC No. | D$^a$ | M$_B^b$ | $B-V^c$ | $\Delta B$ | NGC No. | D$^a$ | M$_B^b$ | $B-V^c$ | $\Delta B$ |
|---|---|---|---|---|---|---|---|---|---|
| 224 | 0.8 | −21.06 | 0.76 | 0.07±0.14 | 4258 | 6.4 | −20.42 | 0.61 | 0.21±0.32 |
| 253 | 2.5 | −19.78 | ⋯ | 0.61±0.44 | 4321 | 17.0 | −21.14 | 0.66 | −0.01±0.26 |
| 300 | 2.1 | −18.06 | 0.56 | −0.02±0.14 | 4559 | 9.7 | −20.13 | 0.35 | −1.38±0.66 |
| 598 | 0.8 | −18.73 | 0.46 | −1.05±0.14 | 4725 | 12.4 | −20.69 | 0.68 | 0.71±0.66 |
| 1566 | 13.4 | −20.46 | 0.57 | −0.87±0.66 | 4736 | 4.3 | −19.40 | 0.73 | −0.07±0.66 |
| 2336 | 33.9 | −22.01 | 0.53 | −0.86±0.66 | 4939 | 44.3 | −22.18 | 0.54 | −1.17±0.66 |
| 2403 | 3.2 | −19.24 | 0.38 | −0.31±0.34 | 5033 | 18.7 | −21.08 | 0.48 | −0.23±0.66 |
| 2541 | 10.6 | −18.46 | 0.36 | −0.28±0.66 | 5055 | 7.2 | −20.19 | 0.67 | 0.47±0.66 |
| 2903 | 6.3 | −19.87 | 0.59 | 0.50±0.66 | 5457 | 7.6 | −21.40 | 0.44 | −1.26±0.25 |
| 2997 | 13.8 | −20.95 | ⋯ | −0.49±0.66 | 6384 | 26.6 | −21.39 | 0.58 | −0.38±0.66 |
| 3031 | 3.6 | −20.38 | 0.86 | 0.64±0.30 | 6946 | 5.5 | −20.82 | 0.40 | −0.24±0.66 |
| 3198 | 10.8 | −19.75 | 0.46 | −0.34±0.66 | 7331 | 14.3 | −21.23 | 0.70 | −0.07±0.66 |
| 3521 | 7.2 | −19.95 | 0.73 | 1.06±0.66 | 7793 | 2.8 | −17.83 | ⋯ | 0.51±0.44 |
| 3621 | 7.1 | −20.01 | 0.47 | −0.43±0.66 | | | | | |

$^a$Distances in Mpc from Tully (1988) based on recessional velocity and Virgocentric infall from Tully & Shaya (1984) except for galaxies with Cepheid distances (Freedman & Madore 1990 (NGC 224); Freedman et al.1992 (NGC 300); Freedman, Wilson, & Madore 1991 (NGC 598); Freedman & Madore 1988 (NGC 2403); Freedman et al.1994a (NGC 3031); Freedman et al.1994b (NGC 4321), Kelson et al.1995 (NGC 5457)) or maser distance (Miyoshi et al.1995 (NGC 4258)).

$^b$Absolute blue magnitudes from Tully (1988), Pierce and Tully (1988), and RC3.

$^c$Total $B-V$ color corrected for internal extinction from RC3.

1) measure the "excess" blue flux from a galaxy and thus the deviation from the mean SFR for galaxies of its mass. The correlation between $B-V$ and $\Delta B$ is significant at more than the 99.99% level (Spearman rank correlation coefficient, $R_S$, of 0.736). The strength of this correlation and of that between H$\alpha$ and $B-V$ (Kennicutt & Kent 1983) indicate that $\Delta B$ is dominated by deviations from the average SFR, rather than by observational uncertainties such as poorly determined extinction corrections. $B-V$, which is distance independent, and $\Delta B$, which is mass independent, form a reliable set of parameters with which to trace recent star formation in this set of nearby galaxies.

The simplest chemical abundance gradient to measure in spiral galaxies is that of O/H based on the bright emission lines of H II regions. Simple empirical calibrations of the ratio ([O II]$\lambda\lambda 3726, 3729$+ [O III]$\lambda\lambda 4959, 5007$)/H$\beta$ provide an estimate of the oxygen abundance relative to hydrogen (cf. Edmunds & Pagel 1984). The details of this calibration are a matter of debate (cf. Henry & Howard 1994), in particular at O/H abundances above solar, but the ranking of abundances appears fairly secure. There are several reasons to suspect that recent galactic evolution might affect $G$: (1) $G$'s are affected by dynamical phenomena such as bars (Pagel et al. 1979; ZKH; Martin & Roy 1994; Friedli, Benz, & Kennicutt 1994), (2) oxygen traces recent star formation because it is produced by Type II SNe, and (3) infalling gas is likely to have a lower metallicity than disk gas.

If O/H correlates strongly with $W_R^i$ at all radii, then $G$ will also correlate with $W_R^i$. O/H as measured roughly midway through the disks of spiral galaxies correlates strongly with $W_R^i$ ($R_S = 0.77$; ZKH), but $G$ does not (ZKH). Therefore, is the observed variation in $G$ among spirals due to unexplained abundance properties at small or large radii? The correlations between $W_R^i$ and O/H, as measured either in the center or at the isophotal radius, have $R_S$'s of 0.72 and 0.37, respectively. The breakdown of the correlation between O/H and $W_R^i$ at large radius indicates that the scatter in $G$ is due principally to abundance anomalies at large radius. The outer disk abundances are more heterogeneous either because the "equilibration" timescale is long or because external factors disturb the abundances. The option that the observed inhomogeneities reflect initial inhomogeneities is unlikely because the gas and stars at the isophotal radius have completed more than ten galactic orbits and existed for many star formation timescales. The latter option, which presumably involves either accretion or interactions, is intriguing.

3. DISCUSSION

Are variations in SFR and $G$ correlated and is the sense of any correlation consistent with the accretion model? The data are shown in Figure 1. The large uncertainties along both axes complicate the issue by *possibly* masking an underlying correlation. The proper line-fitting method for data with large uncertainties along both axes is a maximum likelihood version of linear least-squares (cf. Stetson 1990). That fit is shown in Figure 1. The derived slope, 0.30±0.04, differs from zero at the 7.5$\sigma$ level and has an acceptable $\chi^2 = 1.5$. With the rejection of NGC 2541 from the sample, the only galaxy with a positive $G$, the value of $\chi^2$ falls to 1.2 (and the slope of the fit is now $0.23 \pm 0.04$).

Alternatively, the more robust Spearman rank correlation test indicates that there is only a 3% chance that random data would have as large an $R_S$. However, because of the large uncertainties, the underlying correlation might be significantly tighter. $R_S$ for the eight galaxies with precise velocity-independent distances (highlighted in Figure 1) is much greater than that for the entire sample, 0.76 vs. 0.42, but the random probability of such a correlation is still only about 3% because of the small number of galaxies in this subsample (for comparison, $R_S$ for the same eight galaxies for the correlation between $W_R^i$ and $M_B$, the basis of the TF relationship, is $-0.71$). The improvement in $R_S$ when using only the velocity-independent distances suggests that observational errors are masking the underlying correlation.

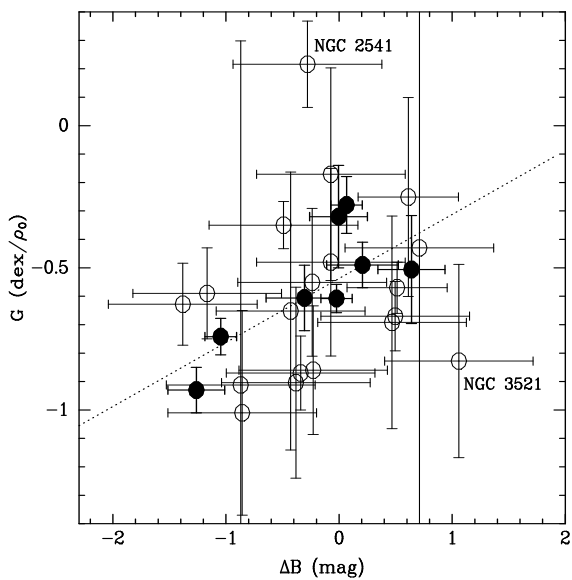

Figure 1: Residuals from the B-band Tully-Fisher relationship are plotted against slope of the O/H abundance gradient. Two galaxies discussed in the text are labeled. Filled circles indicate galaxies for which there are precise velocity-independent distance measurements.

Finally, a new TF relationship is derived using the galaxies with precise distances and $i > 30°$: $M_B = -19.73 - 5.82(\log W_R^i - 2.5)$. The correlation between $G$ vs $\Delta B$ now has $R_S = 0.42$ (vs. 0.42 previously) and a maximum likelihood least-squares slope of $0.27 \pm 0.04$ with $\chi^2 = 1.4$ (an $R_S$ and slope of 0.41 and $0.20 \pm 0.04$ without NGC 2541). Therefore, the results are insensitive to the detailed choice of TF relationship.

A maximum likelihood estimate of the intrinsic scatter in the underlying correlation was obtained by simulating data that match the distribution of observed $\Delta B$'s, lie along the line derived from the maximum likelihood fit, and have an intrinsic scatter $\sigma$ about that line. For a choice of $\sigma$, observational uncertainties are then added to the simulated data and $R_S$ is evaluated. For $\sigma = 0.12$ mag, one-half of the simulated samples have $R_S$ less than the observed value and the underlying distribution has $R_S = 0.84$ (random probability $< 10^{-7}$). These results further indicate that the correlation between $G$ and $\Delta B$ is real.

As discussed above, an alternative SFR indicator is $B-V$. The trend between $G$ and $B-V$ is similar to that between $G$ and $\Delta B$, but is not statistically significant when all of the galaxies for which $B-V$ measurements exist are used. By removing NGC 2541 from the sample the likelihood of the correlation arising at random drops to 2% and the correlation is in the expected sense (that bluer galaxies have steeper gradients). Recall that $B-V$ and $\Delta B$ have different weaknesses, so the agreement is encouraging although not unexpected due to the strong correlation between $B-V$ and $\Delta B$.

So far I have presumed that a change in $G$ reflects solely a change in the abundance properties rather than a change in the length scale over which $G$ is measured. An interaction between a companion and parent galaxy might change the parent galaxy's isophotal size and so indirectly change $G$. Although this effect does not dominate — the correlation between the isophotal radius expressed in physical units and $\Delta B$ is not significant — there is a trend that "bluer" galaxies are physically larger.

This trend may be related to the final prediction of the accretion model: systems with recent accretion should exhibit some morphological signature. The simplest signature is an asymmetry at large radius where the accretion occurred. Two of the most striking examples of H I asymmetries are in NGC 628 (Briggs et al. 1980) and NGC 5457 (M101; Bosma, Goss, & Allen 1981). As expected in the accretion scenario, these galaxies have highly negative $G$'s (they are among the five galaxies with steepest gradients, although NGC 628 is not included in Table 1 because $i < 30°$) and M101 has the second most negative (blue) $\Delta B$. In addition, Bosma, Goss, & Allen note that all of the H II regions at large radius in M101 are in the asymmetric extension, which would further account for the low oxygen abundances because some of that material is presumably from a low-mass, and hence low metallicity (Skillman, Kennicutt, & Hodge 1989), companion. NGC 3344, another one of the five galaxies with steepest gradients, also has an extended H I distribution (Corbelli, Schneider, & Salpeter 1989), although only a low resolution map is available. Because high-sensitivity high-resolution maps for all of the galaxies in the sample are unavailable, it is impossible to quantitatively and systematically test the connection between H I asymmetries and star formation/abundance gradients. The lifetime of the H I asymmetries in NGC 628 or M101 depends on details of the gravitational potential. However, simple winding arguments suggest a lifetime of several Gyr for standard parameters. If so, then there is sufficient time for star formation in the gas asymmetry to generate a corresponding stellar asymmetry.

The only systematic and quantitative compilation of asymmetries in stellar disks is that based on the K-band ($2.2\mu$m) observations of Rix & Zaritsky (1995; hereafter RZ). Unfortunately, that sample contains no galaxies with measured $G$'s, so the only possible test is for a correlation between disk asymmetry and SFR. Their sample consists of face-on spiral galaxies, so I will use $B-V$, rather than $\Delta B$, as the SFR indicator. $(B-V)_T^0$'s from RC3 exist for only 12 of the 18 galaxies in their sample. $R_S$ for the correlation between $B-V$ and the stellar disk asymmetry at 2.5 disk scale lengths (cf. Table 2 of RZ) is $-0.389$. This value suggests a correlation in the expected sense (i.e., bluer galaxies are more asymmetric), but the correlation is not statistically significant. There is no correlation between asymmetry and $M_B$, which indicates that the sample is not biased by blue, low-mass, irregular galaxies.

Such arguments can provide a measure of the accretion rate, although more thorough calculations regarding the lifetime of the asymmetries, and more observations, must be done before such results can be fully accepted. Nevertheless, the appearance of stellar asymmetries, which RZ estimate last about 1 Gyr, in about 1/3 of field spirals and the equal numbers of galaxies with high or low $G$ suggest that accretion events in field spiral galaxies were common over the last few Gyr.

## 4. SUMMARY

The data presented here *suggest* that relatively isolated spiral galaxies have experienced accretion of companion galaxies or clouds over the last few Gyr. Those familiar with the situation in our own Galaxy may not find this conclusion surprising. The estimate of the orbital decay time of the LMC (Tremaine 1976), the existence of the Magellanic Stream (Lin & Lynden-Bell 1977), and the tidal disruption of the Sagittarius dwarf (Ibata, Gilmore, & Irwin 1994; Johnston, Spergel, & Hernquist 1995) all suggest that our galaxy has and will experience accretion events. Accretion, or more appropriately the lack of accretion, may also explain why Virgo galaxies with stripped H I envelopes have larger mean chemical abundances and flatter abundance gradients than normal galaxies (Skillman *et al.* 1995). The goal of the work presented here is to identify some observational signatures of recent accretion.

The results from this study are that (1) the variation in abundance gradient slopes ($G$) for galaxies is principally due to differences in the outer disk abundances, (2) the residuals from the B-band Tully-Fisher (TF) relationship correlate strongly with $B - V$, indicating that the residuals from the B-band TF relationship reflect different levels of the mass-normalized SFR activity among galaxies, (3) $G$ correlates with both $B - V$ and B-band TF residuals, in the sense that galaxies with steeper gradients have an elevated SFR, and (4) there is anecdotal evidence for large H I asymmetries in galaxies with steep abundance gradients and larger than average SFR's (NGC 628 and M101 are the two best examples) and a suggestion that spiral galaxies with asymmetric stellar disks have elevated SFRs. These results are all consistent with a simple accretion model, although not exclusive of other scenarios.

The interactions between a parent galaxy and its companion suggested here are mild accretion events and probably do not even qualify as "minor mergers" ($M_{comp} = 0.1 M_{par}$). Simulations suggest that the gas in minor mergers is quickly driven to the nucleus, especially in bulge-poor parents (Mihos & Hernquist 1995). The end-products of such mergers are unlikely to be included in a sample of "normal" late-type spirals such as that presented here.

While it is generally accepted that galaxies grow and evolve hierarchically, little is known about the rate and effects of the accretion events. Based on measurements of $G$'s, SFR's, and asymmetric gas and star distributions, I conclude that there is evidence for relatively recent ($\sim$ few Gyr) accretion events in 1/3 to 1/2 of nearby, relatively isolated, spiral galaxies and that the aftermath of these events is observable. The samples on which these conclusions are based are small but will soon improve. Only eight of the galaxies in Table 1 have precise distances ($\Delta D/D < 0.15$). However, many of the other galaxies are targets of the Cepheid HST Distance Scale key project (NGC 2541, NGC 3198, NGC 3351, NGC 3621, NGC 4725, NGC 7331) and NGC 7793 is being studied from the ground. Precise distances should be available within the next few years, which should greatly improve our understanding of the correlations suggested by the present data. Additional observations of H II regions and H I at large radius will also be key in probing the effects of accretion on galaxy evolution.